\documentclass[aps,prb,showpacs,twocolumn]{revtex4}
\usepackage{epsfig}
\usepackage{amsmath}

\begin{document}

\title{Disorder effect on the Friedel oscillations in a one-dimensional Mott insulator}
\author{Y. Weiss, M. Goldstein and R. Berkovits}
\affiliation{The Minerva Center, Department of Physics, Bar-Ilan University,
  Ramat-Gan 52900, Israel}

\begin{abstract}

The Friedel oscillations resulting from coupling a quantum dot to one edge 
of a disordered one-dimensional wire in the Mott insulator regime, are calculated numerically
using the DMRG method. By investigating the influence of the disorder on the 
Friedel oscillations decay
we find that the effect of disorder is reduced by increasing the interaction strength.
This behavior is
opposite to the recently reported influence of disorder in the Anderson insulator regime,
where disorder led to a stronger decay of the Friedel oscillations.

\end{abstract}


\maketitle

Properties of one-dimensional systems incorporating disorder and electron-electron 
interactions are the subject of many recent theoretical and
experimental studies. 
It is well known that when the interactions are not too strong, the addition of 
disorder turns the metallic system into an Anderson insulator (AI). However, for strong 
interactions (i.e., when the clean system is a Mott insulator) the exact 
effect of disorder depends on its strength, and in general is not completely understood.
While for clean systems the Mott insulator (MI) phase is a well studied problem\cite{mott90},
the addition of disorder opens a few questions, which have attracted
several studies in the last decade\cite{pang93,fujimoto96,mori96,orignac99,giamarchi01}.

Specifically, since most studies on disordered one-dimensional wires concentrate on 
either the AI or the MI phases, a full comparison between the two regimes is still lacking. 
Nevertheless, a qualitatively different behavior between these two regimes was 
demonstrated in a few cases. For example, the 
effect of interactions on the persistent current in one-dimensional disordered rings
was calculated in previous works \cite{abraham93,bouzerar94}, and an important
difference between the AI and MI phases was found. 
While for strong interactions and weak disorder (MI phase) 
the persistent current was reduced, for strong disorder (AI phase) an increase
of the current was found. However, the exact diagonalization techniques which 
were used in these studies, are applicable only for very small system sizes.

The definition of a disordered MI phase should be refined, since
when the disorder is strong, i.e., when the random potential felt by the electron 
is much larger than any other energy scale in the problem, the MI state is destroyed.
For a weak disorder, however, it was shown in several studies that 
the Mott energy gap vanishes only when a finite disorder is introduced, so that
below this critical disorder the MI phase is stable \cite{ma,sandvik94}. 
Usually this is not the case for a MI consisting of spinless particles, 
since an Imry-Ma type of argument \cite{imry_ma} 
shows that the long range order is destroyed even for an infinitesimal disorder \cite{shankar}. 
Yet, for a finite sized mesoscopic sample, the Imry-Ma length scale might 
be a few orders of magnitude larger than the sample's size,
so that the effective ground state for a weak enough disorder remains a MI one.
Such finite one dimensional wires coupled to dots have been recently
manufactured, and signatures of a charge density wave in strong magnetic
fields have been observed \cite{pepper}. 
Increasing the disorder above a critical strength changes the MI state either
to a Mott Glass or to an AI \cite{orignac99,giamarchi01}. 

In this paper we investigate the influence of interactions on the Friedel oscillations (FO) 
in a disordered one-dimensional wire, and compare this behavior between the AI and the MI regimes.
We study interacting spinless electrons confined to a 1D wire which can be in either 
its AI or MI phases. Without disorder, it is known that in order to get a MI phase 
the repulsive e-e interactions should be strong enough, while for weaker interactions 
the wire is described by the Tomonaga-Luttinger liquid (TLL) theory.
The MI phase, for strong interactions, appears for spinless 1D electrons as a $2k_F$ 
charge density wave (CDW). 
When disorder is included, the TLL phase switches into an AI state.
However, the finite size CDW state is expected, as noted above, to remain 
stable against the application of a weak enough disorder, i.e. to remain a MI state. 
For example, previous numerical simulations have presented the long range order of 
such a weakly disordered CDW
\cite{pang93}. 
In order to verify the existence of the CDW order in the presence of disorder 
for the length scales considered, one should check the electron density of 
the entire system.

The behavior of the FO decay length in the presence of
disorder in the AI phase,
were discussed in a recent paper \cite{ours_FO_LL}.
It was shown that the effect of disorder on the FO decay length can be described by an 
additional exponential term $e^{-x/\xi}$, where $\xi$ is a characteristic decay length.
For a constant strength of (weak) disorder $\xi$ decreases as the 
interactions increase, i.e. the disorder effect is enhanced with increasing interactions.
Arguing that $\xi$ is a good approximation of the mobility localization length,
it was found that it is in good accordance with theoretical predictions made 
using the TLL framework\cite{apel_82,suzumura_83,giamarchi}, 
which are suitable for the weak interactions regime. 
However, for the CDW phase umklapp processes are important, and the above
considerations are not applicable.

In order to calculate the decay length of the weakly disordered CDW wire, 
we use a method similar to the one used for the TLL regime. We couple the wire to a 
quantum dot with a single level from one end, and 
study the electrons density change in the sites nearby. The density change, which have shown
Friedel oscillations with a $2k_F$ wave vector and a power law decay in the metallic case (TLL), 
should now present $2k_F$ oscillations with an exponential decay, since the underlying 
lattice state (CDW) is an insulator. 
For the clean case we will show that the exponential decay length scales as 
the CDW correlation length, $\zeta$, as predicted \cite{mikeska}. However, in the disordered case we 
find an additional decay factor due to the disorder, as in the TLL case\cite{ours_FO_LL}. 
By calculating this decay length we are able to present a clear picture of the dependence 
of the decay length due to disorder on interactions, in both the AI and MI regimes.
While the decay length of the FO due to disorder 
is monotonically decreasing as interaction increases for the AI phase, for the MI phase 
it is monotonically increasing. The origin of the difference between these two 
regimes will be explained.

The system under investigation is the strong electron electron interactions regime of a 
one-dimensional wire with disorder. The wire is modeled by a one-dimensional lattice
of spinless fermions, moving in a random on-site potential, and experiencing nearest 
neighbor repulsive interactions. The Hamiltonian is
\begin{eqnarray} \label{eqn:H_wire}
{\hat H_{wire}} = 
\displaystyle \sum_{j=1}^{L} \epsilon_j {\hat c}^{\dagger}_{j}{\hat c}_{j}
-t \displaystyle \sum_{j=1}^{L-1}({\hat c}^{\dagger}_{j}{\hat c}_{j+1} + h.c.) \\ \nonumber
+I \displaystyle \sum_{j=1}^{L-1}({\hat c}^{\dagger}_{j}{\hat c}_{j} - \frac{1}{2})
({\hat c}^{\dagger}_{j+1}{\hat c}_{j+1} - \frac{1}{2}),
\end{eqnarray}
where $\epsilon_j$ are the random on-site energies, taken from a uniform 
distribution in the range $[-W/2,W/2]$,
$I$ is the nearest neighbor interaction strength, and $t$, which is the
hopping matrix element between nearest neighbors, sets the energy unit scale.
${\hat c}_j^{\dagger}$ (${\hat c}_j$) is the creation (annihilation)
operator of a spinless electron at site $j$ in the wire, and
a positive background is included in the interaction term.
Such a (clean) wire undergoes a phase transition at $I=2t$ between TLL and CDW.
In order to study the CDW and the weakly disordered CDW phase the interaction strength is taken 
to be strong, i.e. $I>2t$.

We now introduce a quantum dot with a single orbital at one end of the wire, by adding 
the following term to the Hamiltonian:
\begin{eqnarray} \label{eqn:H_imp}
{\hat H_{dot}} = \epsilon_0 {\hat c}^{\dagger}_{0}{\hat c}_{0} 
-V ({\hat c}^{\dagger}_{0}{\hat c}_{1} + h.c.) \\ \nonumber
+I ({\hat c}^{\dagger}_{0}{\hat c}_{0} - \frac{1}{2})
({\hat c}^{\dagger}_{1}{\hat c}_{1} - \frac{1}{2}),
\end{eqnarray}
where $\epsilon_0$ describes the dot energy level.
As in Ref.~\onlinecite{ours_FO_LL}, we take $\epsilon_0 \gg W$ and $V=t$.

The Hamiltonian $\hat H$ was diagonalized using a finite-size DMRG method
\cite{ours_FO_LL,white93}, and the occupation of the lattice sites were calculated,
for different values of $W$ and $I$. The dot energy level was taken to be $\epsilon_0=10t$.
The size of the wire was up to $L=300$ sites,
which is both long enough due to the exponential decay of the calculated quantities,
and still short enough to maintain the CDW order for the disorder strengths taken ($W/t=0.1$ and $0.2$).
During the renormalization process the number of particles in the system is not fixed, 
so that the results describe the experimentally realizable situation of a
finite section of a 1D wire which is coupled to a dot and to an external electron 
reservoir\cite{ours2}.
Yet, the calculated density remains close to half filling in all the calculated
scenarios (even in the presence of disorder)
since the interaction term contains a positive background, 
and the calculation is done for $\mu=0$.

We start with the case in which
no disorder is considered ($W=0$), so that the ground state of the CDW is
twofold degenerate. This degeneracy is broken, however, once a pinning impurity,
denoted by $\epsilon^{(0)}_0 \rightarrow 0^+$, is coupled to one end of the wire, and the wire shows
a $2k_F$ modulation \cite{ours2}. The particle density of such a state, 
in the $j$-th site of the wire, will be denoted by $N^0_j$. 
When the pinning impurity is replaced by a dot level with $\epsilon_0 \gg \epsilon^{(0)}_0$, 
the particle density in the wire (to be denoted as $N_j$) is changed by an oscillating
$2k_F$ term. 
$2k_F$ oscillations in the density difference were also obtained in the TLL phase
(Ref.~\onlinecite{ours_FO_LL}), where the density without the quantum dot is flat, 
and the deviation of the population from this flat density
once the lead is coupled to the dot shows Friedel oscillations.
Here one should notice that the reference state (without the dot) does not have a 
flat particle distribution, but rather has a CDW $2k_F$ oscillations. Coupling the dot 
results in a new CDW state, which has also $2k_F$ oscillations, but with a different
amplitude. The difference between these two states has a  
$2k_F$ oscillation, which has an exponential decay from its value at the edge of
the wire. 

\begin{figure}[htb]\centering
\vskip 0.6truecm
\epsfxsize8.0cm
\epsfbox{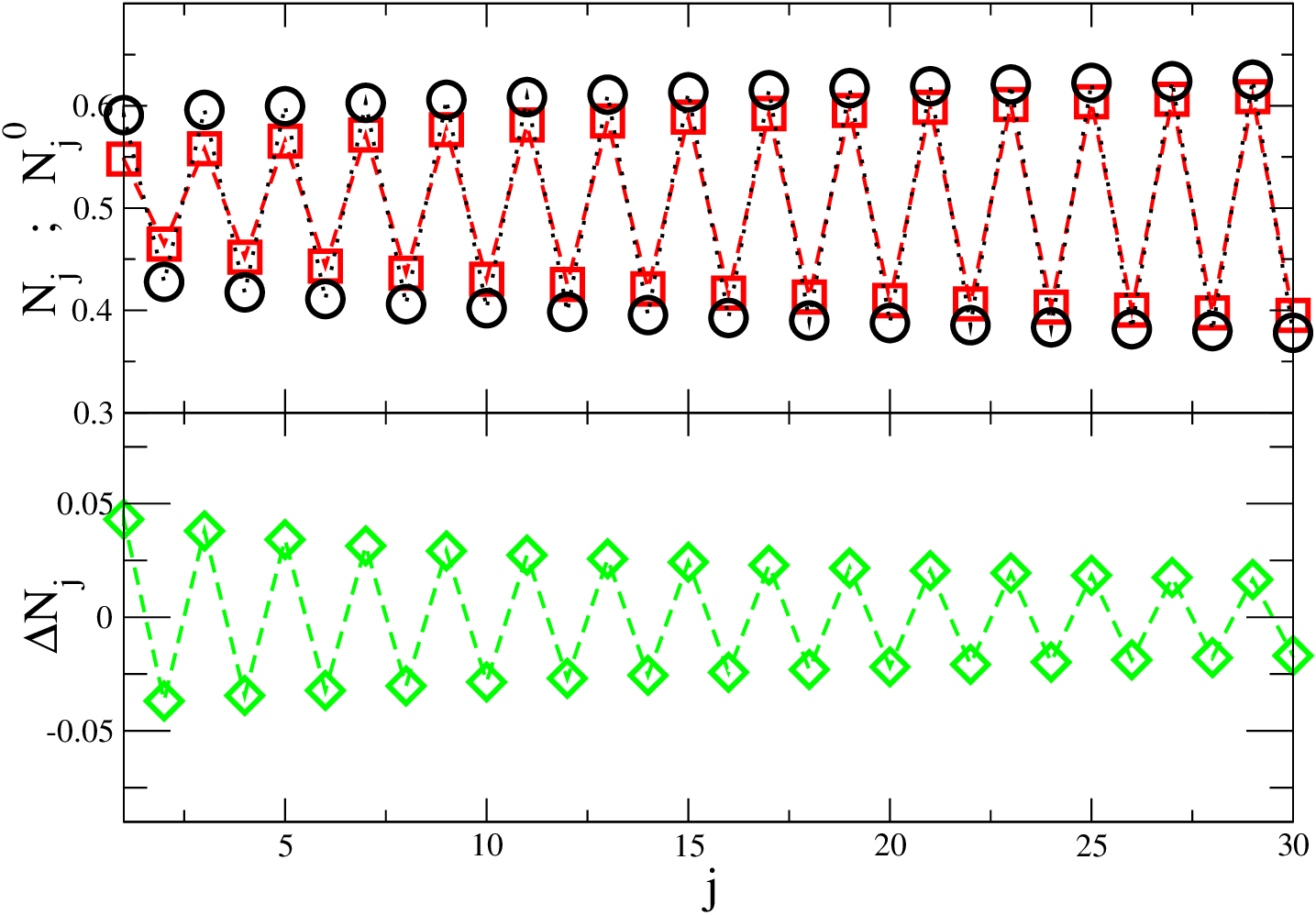}
\caption{\label{fig:fo_sample_W0}
Typical oscillations for a clean sample with $L=280$ 
for a CDW with $I/t=2.5$. 
The upper panel shows $N_j$ (circles) and $N^0_j$ (squares),
and the lower panel presents their difference $\Delta N_j$.
}
\end{figure}

In order to calculate the density difference between the cases when the
quantum dot is coupled or uncoupled to the wire, 
one defines the density change in site $j$ as 
\begin{eqnarray} \label{eqn:Y_j_1}
\Delta N_j = N_j - N^0_j,
\end{eqnarray}
and studies the dependence of $\Delta N_j$ on $j$ for different parameters.
A typical result of $N_j$ vs. $N^0_j$, and the resulting $\Delta N_j$, 
showing the $2k_F$ oscillations caused by the dot orbital at $j=0$,
is presented in Fig.~\ref{fig:fo_sample_W0}. 

When $W \ne 0$, the CDW ground state is no longer degenerate, and the infinitesimal pinning impurity is
not required. The disorder itself pins the CDW to different places on the lattice,
with the ability to break the long range order of the clean CDW by localized solitons, 
with a density which depends on the disorder strength\cite{pang93}.
Yet, when a dot level with $\epsilon_0 \gg W$ is connected to one side 
of the wire, the local effect in its vicinity is stronger than the 
pinning caused by the disorder. This results in a change of the particle 
density near the dot, and this change decreases with distance. It 
turns out that the definition of $\Delta N_j$ in Eq.~(\ref{eqn:Y_j_1})
is suitable for the disordered case as well, since it cancels out the 
disorder pinning effects which are the same for the two cases, isolating 
the density fluctuations created by the dot.

\begin{figure}[htb]\centering
\vskip 0.6truecm
\epsfxsize8.0cm
\epsfbox{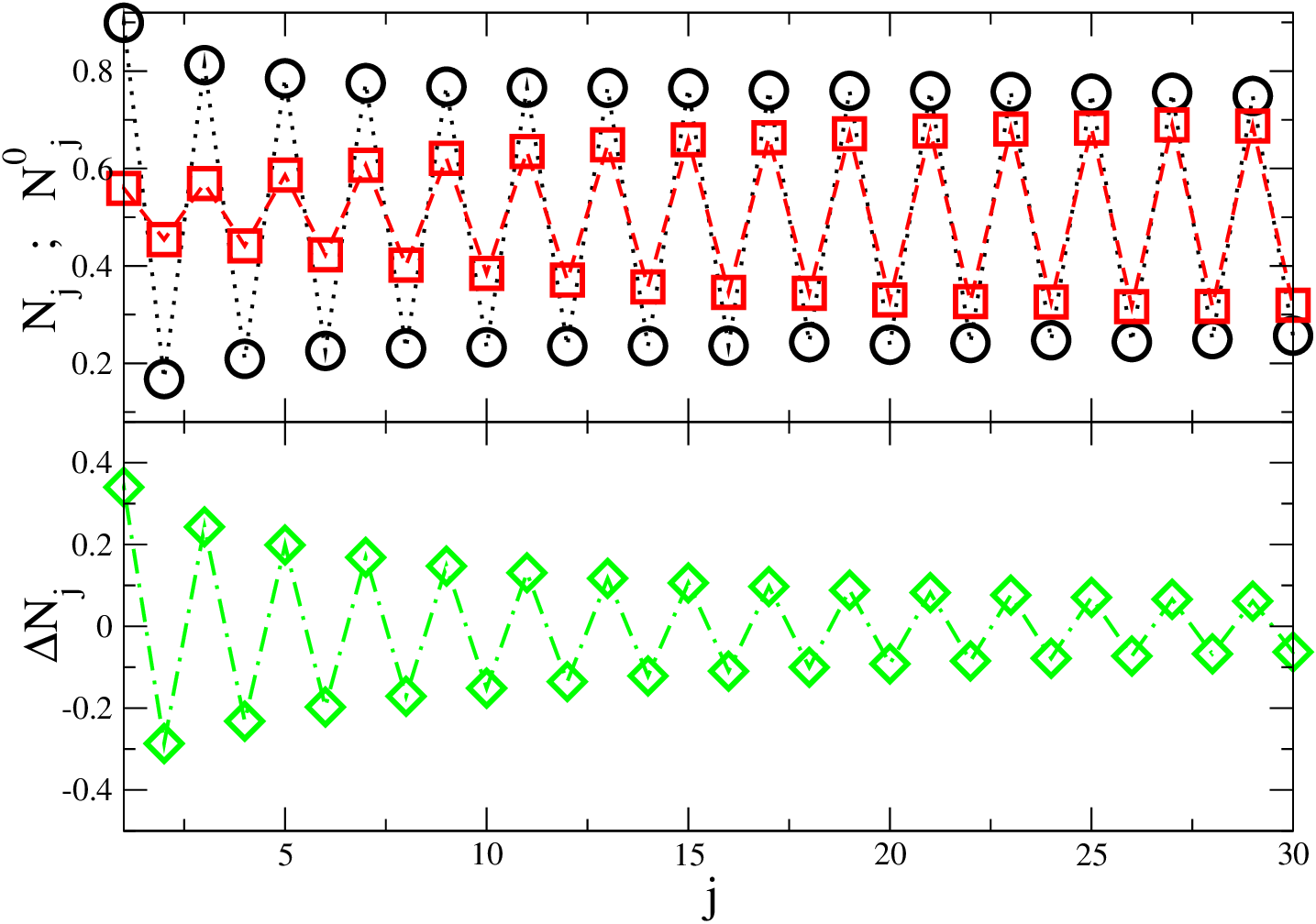}
\caption{\label{fig:CDW_frid_sample}
Typical oscillations for a single disordered sample with $L=280$, $W=0.1$ and $\epsilon_0=10$, 
for a CDW with $I=3$. 
The upper panel shows $N_j$ (circles) and $N^0_j$ (squares),
and the lower panel presents their difference $\Delta N_j$.
}
\end{figure}

A typical picture of $\Delta N_j$ for a disordered CDW sample is presented in 
Fig.~\ref{fig:CDW_frid_sample}. Whereas the upper panel shows the density of the
two similar systems, one which is coupled to the quantum dot and the other is not, the lower panel presents
the difference between these two densities, and the decay of the oscillations can be clearly seen.

\begin{figure}[htb]\centering
\vskip 1.0truecm
\epsfxsize8.0cm
\epsfbox{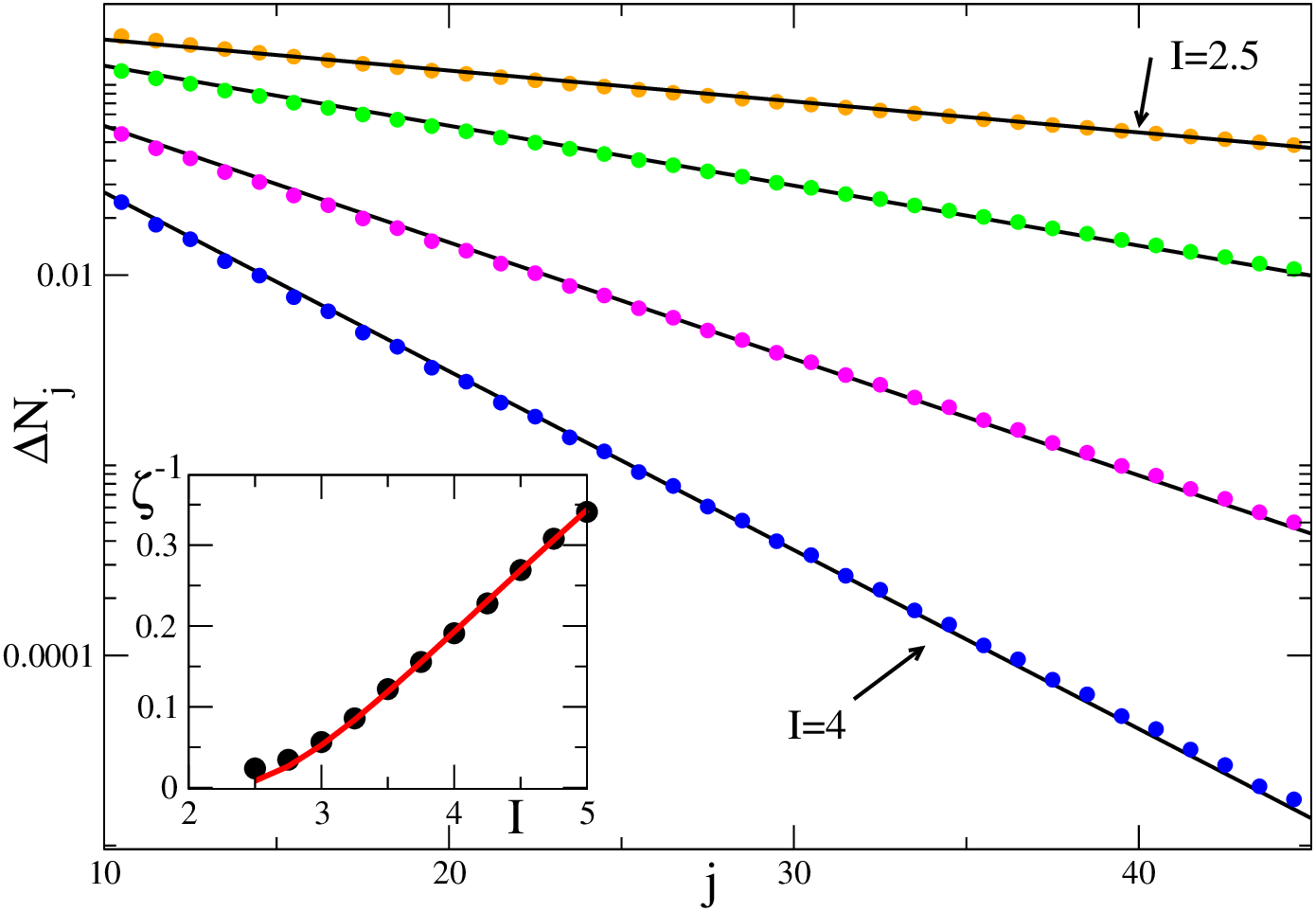}
\caption{\label{fig:cdw_frid}
The oscillations decay in the CDW regime for various interaction strengths and without disorder
(note the semi-log scale).
As the interaction increases, the correlation length decreases and the decay is faster.
Inset: the inverse correlation length of the CDW state for various interaction strengths (symbols)
fitted to the theory prediction Eq.~(\ref{eqn:zeta}).
}
\end{figure}

Since the CDW is an insulating phase, the decay of the $2k_F$ oscillations without disorder is supposed to 
be exponential and the characteristic length is the CDW correlation length \cite{pang93},
i.e., $\propto \exp(-x/\zeta)$. 
In Fig.~\ref{fig:cdw_frid} such an exponential 
decay of $\Delta N_j$ is shown on a semi-log scale for various interaction strengths.
An exact Bethe Ansatz solution \cite{mikeska} of our model gives the relation between 
the correlation length and the interaction as
\begin{eqnarray} \label{eqn:zeta}
\zeta \sim exp(\pi/ \sqrt{I/(2t) - 1}).
\end{eqnarray}

The correlation lengths extracted from
the DMRG results are presented with a fit to the exact formula in the inset of Fig.~\ref{fig:cdw_frid}.
As can be seen, for $I$ not very close to the TLL-CDW transition point (which occurs at $I=2t$), 
the results fit the theory very well.

\begin{figure}[htb]\centering
\vskip 0.6truecm
\epsfxsize8.0cm
\epsfbox{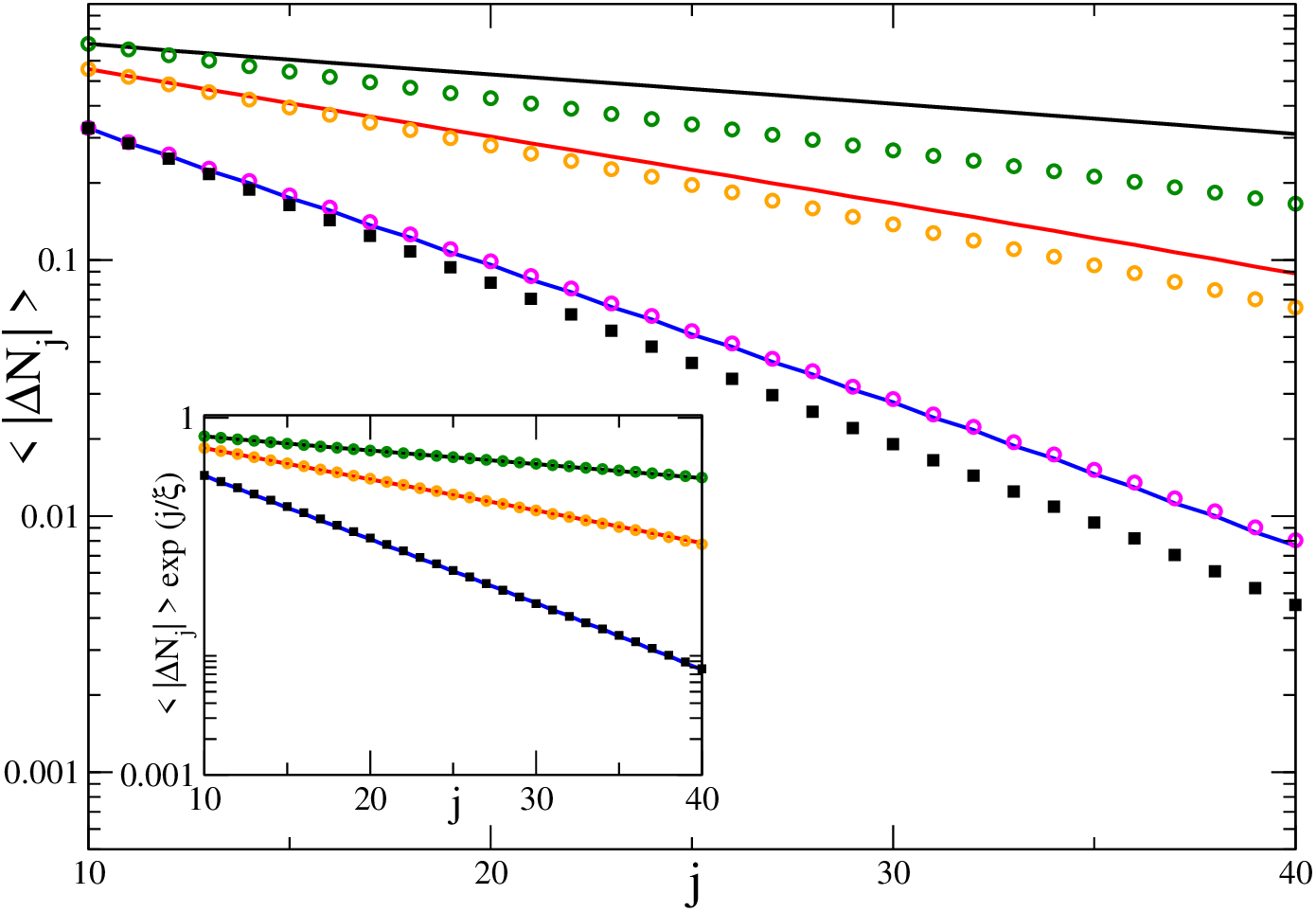}
\caption{\label{fig:CDW_decay_I3.5}
The decay of the oscillations of a disordered CDW with $I=2.5,3$ and $3.5$ (top to bottom,  
note the semi-log scale). 
The lines correspond to the clean sample result, and the symbols to the averaged 
disordered data. For $W=0.1$ (circles) the disorder effect 
is clearly seen for $I=2.5$ and $I=3$ but not for $I=3.5$ in which $\xi$ is
much larger than the correlation length $\zeta$. For $W=0.2$ (squares) $\xi$ is small 
enough to affect the decay even for $I=3.5$.
Inset: multiplying $\Delta N_j$ by $e^{x/\xi}$ collapses the disordered data on the clean curves.
}
\end{figure}

For $W \ne 0$, $\Delta N_j$ is averaged over $100$ realizations,
for which we expect a sampling error of the order of one percent. 
Assuming that the disorder adds another exponential term to the oscillations decay,
which is thus proportional to $\exp(-x/\zeta-x/\xi)$,
there are two competing length scales - the decay length due to disorder ($\xi$) vs. the correlation length ($\zeta$). 
For strong interactions and weak disorder $\zeta \ll \xi$ so 
that the disorder effect is hardly seen, but 
increasing the disorder or decreasing the interaction strength 
should result in a combination of the two exponential decays.
The DMRG results, presented in Fig.~\ref{fig:CDW_decay_I3.5}, show the disorder 
effect on the oscillations decay. For $I=2.5$ and $I=3$ one can see faster decay for the disordered samples with $W=0.1$.
For stronger interaction larger disorder is required in order to affect the decay.

\begin{figure}[htb]\centering
\vskip 0.6truecm
\epsfxsize8.0cm
\epsfbox{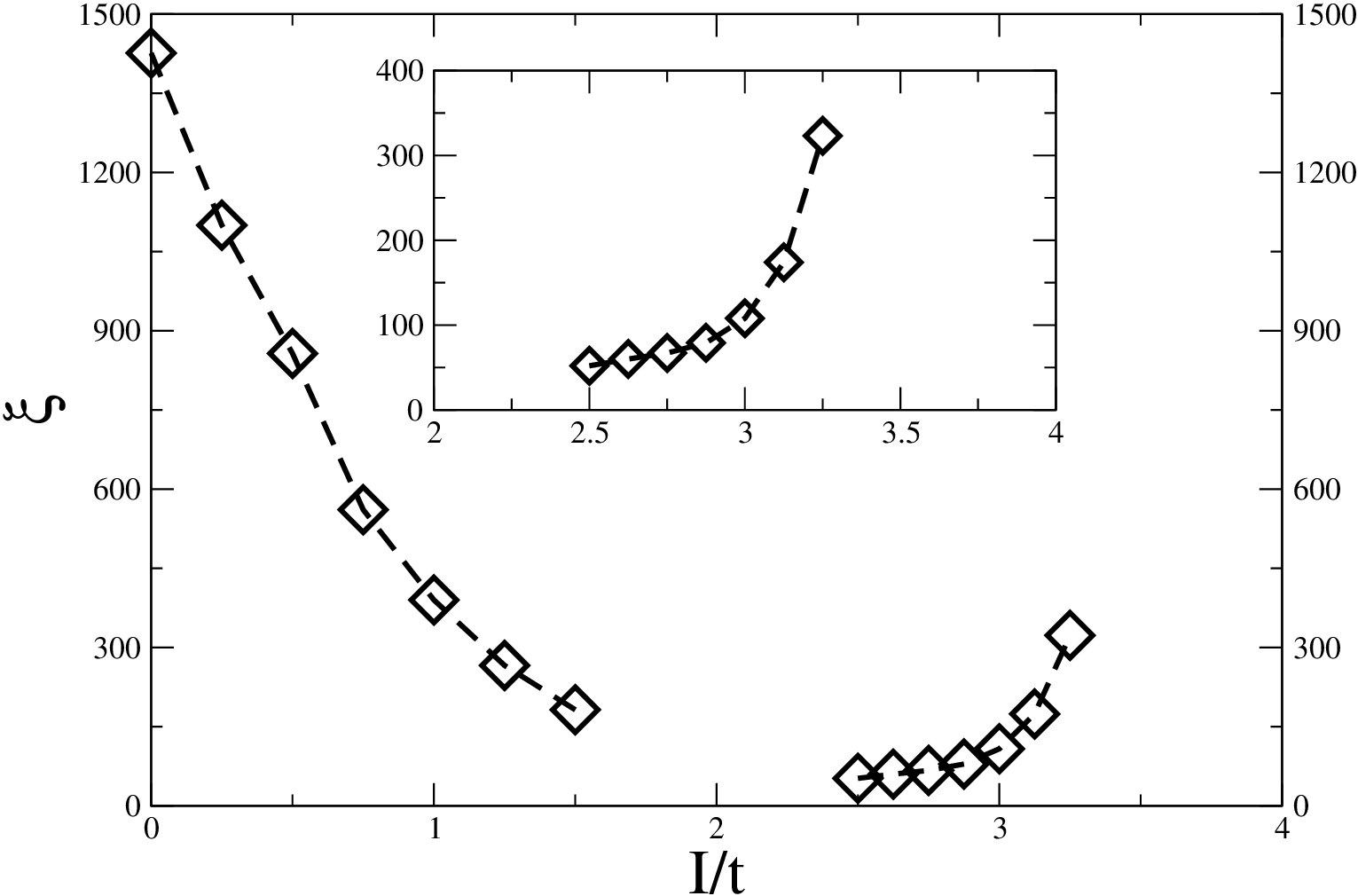}
\caption{\label{fig:xi_I_CDW}
The decay length due to disorder ($\xi$) in the TLL ($I<2t$) and in the CDW ($I>2t$)
phases as a function of the interaction strength. The data for the TLL 
phase was taken from Ref.~\onlinecite{ours_FO_LL}.
Inset: zoom into the CDW regime.
}
\end{figure}

Similarly to the AI phase, the extra decay length can be extracted by fitting, 
for each value of $I$,
the $W \ne 0$ curve multiplied by $e^{x/\xi}$ to the $W=0$ one. Such a 
rescaling is presented in the inset of Fig.~\ref{fig:CDW_decay_I3.5}. 

As can be seen in Fig.~\ref{fig:xi_I_CDW}, 
the decay length extracted for the disordered MI regime increases as
a function of the interaction strength (for $2t < I \lesssim 3.5t$), 
an opposite behavior to the AI case ($I<2t$).
Results for stronger values of $I$ are not shown, since
for too strong interactions the correlation length is very small,
and thus the estimate of $\xi$ is less accurate.

These results point out that as the interaction strength increases in the MI phase, the 
disorder effect decreases. In the AI phase, on the other hand, the disorder effect
is enhanced with increasing interactions.
The difference between these two behaviors results from the difference in the
ground states of the two phases in the clean case. In our model 
there is a competition between the kinetic energy (the hopping term) and the potential
(the interaction). The hopping term prefers the existence of a flat particle
distribution whereas the interaction term prefers a CDW-like form. 
For different values of $I$ the results of that competition are different: for $I<2t$
(the TLL phase) the hopping term wins, and the distribution is flat, while for $I>2t$ 
(the CDW phase) a CDW starts to form.

Inside the clean TLL phase, as $I$ increases, the CDW fluctuations are stronger. Yet, the 
average density profile in the ground state remains flat because of the 
hopping term. But when disorder is introduced, 
the flat density state becomes less favorable than a state with a fluctuating density,
the latter being preferred by both the disorder and the interactions. For a 
constant disorder, as the interactions become stronger, these fluctuations are 
enhanced, so the disorder effect increases.

In the CDW phase, on the other hand, without disorder, the interaction wins over 
the hopping, and the ground state has a CDW form. 
Turning on the disorder might change the particle distribution, e.g. by allowing 
an electron to move into a site with lower on-site energy, but this results in raising 
the interaction energy. As the interaction strength gets stronger, the probability 
of such a process decreases, so that the actual effect of the disorder is getting weaker.

In conclusion, while the decay length of the $2k_F$ oscillations envelope due to disorder 
is monotonically decreasing in the AI phase, we have shown that it is monotonically 
increasing in the disordered MI phase. The difference between these two regimes is explained by
the difference between the ground states of the clean samples in each case. In 
the AI phase the pure ground state is flat, and both the disorder and the interactions try to introduce fluctuations in it. 
In the MI phase, on the other hand, the pure ground state oscillates with a $2k_F$ 
wave vector, and these oscillations are enhanced by the interactions and reduced by the disorder.
As a result, the disorder effect (for a constant disorder strength) is getting weaker as
the interactions are enhanced.

\acknowledgments

Support from the Israel Academy of Science 
(Grant 877/04) is gratefully acknowledged.

\end{document}